# STATISICAL AND MULTIFRACTAL PROPERTIES OF TIME SERIES GENERATED BY A MODIFIED MINORITY GAME

Yu.A Kuperin[*], M.M.Morozova[+]
Saint Petersburg State University
[*] Department of Physics, Saint Petersburg State University, #3 Ulyanovskaya Str., Saint Petersburg, 198054, Russia
E-mail: yuri.kuperin@gmail.com

[+] Smolny College of Liberal Arts and Sciences , Saint Petersburg State University, #56-60 Galernaya Str., Saint Petersburg, 199034, Russia
E-mail: 0minus273@gmail.com

**ABSTRACT**
In this paper it was developed a modification of the known multiagent model Minority Game, designed to simulate the behavior of traders in financial markets and the resulting price dynamics on the abstract resource. The model was implemented in the form of software. The modified version of Minority Game was investigated with the aim of reproducing the basic properties of real financial time series. It was proved that such properties as the clustering of volatility, the Levy distribution and multifractality are inherent for generated by this version of the Minority Game time series of prices.

## 1. INTRODUCTION

The main objective of this work was to develop modifications of the known multi-agent models Minority Game [9, 10, 23], which is designed to simulate the behavior of traders in the financial market subject to the conditions most close to reality. Based on the dynamics of traders such game should generate a time series of prices for an abstract resource, which should have the most important features of real financial time series, namely: multifractality, heavy-tailed probability distributions and volatility clustering [17, 21, 30].

For achieving the above stated objectives there have been addressed several subtasks. First, we have developed a minimal list of modifications of the Minority Game, which was enough to ensure that the modified game would generate a time series with the required properties. Secondly, the software for implementation of Minority Game has been developed. Thirdly, there have been selected or developed methods for multifractal and statistical analysis of the obtained data. These methods have been also implemented in the form of software.

Attempts to modify the original version of the Minority Game in order to obtain multifractal time series have been made before. For example, in [12] the authors have attempted to develop a modification of the Minority Game, which serve to demonstrate the multifractal properties. To achieve this effect, the author had to make not a realistic step: they broke all the players on the two conventional groups - speculators and producers. Moreover, the second group was deprived of the opportunity to skip the game, and could join the bidding only every fifth day (Market Week). Despite the fact that the obtained in [12] time series, actually possess the desired property of multifractality, the question arises whether it is possible to achieve the same results with a more realistic and simple assumptions about the structure and dynamics of the agents. These assumptions and their implementation are presented in this paper.

In the paper [14] was carried out large-scale study of the methodology of conducting the games directly to real financial data. The aim of this work was the prediction of financial time series. This game was also carried out based on the Minority Game. Each agent had two sets of strategies with a fixed memory. The decision which of the sets of strategies used was taken either by a coin flip, either by scoring strategies. Interestingly, the parameter N (number of players) was removed from the game setting $N \to \infty$. Thus, this paper mainly studied the probability distribution when one of the sets of strategies was chosen. Despite the interesting findings, this study was not designed to explain the dynamics of financial markets or to build a good model. This study focused exclusively on the successful prediction and some interesting aspects for our work have not been considered.

In the paper [9] the authors also used a modified Minority Game to simulate financial markets. These authors obtained the following results: heavy tails of the distribution and clustering of volatility, as well as the scaling behavior. But the division of agents in the group was even more complicated than described in [12]: in addition to producers and speculators, they also added noisy traders whose behavior, roughly speaking, was accidental. This is not consistent with the requirements of rational agents, which was the basis of modification, represented in the present paper.

The structure of the article is as follows. The second section provides the necessary information from the fractal and multifractal analysis. In the third section, we describe a new approach to the calculation of the modified point wise Holder exponents (MPHE), first proposed in [19]. The fourth section describes the statistical quantities and methods that are used for the analysis of time series generated by a modified Minority Game. The fifth section contains details about those modifications of Minority Game, which had been made in comparison with its classical version. The sixth section contains the results and their analysis.

## 2. MULTIFRACTAL ANALYSIS

Multifractal analysis [9] is relatively recent mathematical field of study. Due to its universality, this approach is not only effectively used in the field of mathematics or physics. There are many branches of science such as linguistics, medicine, informatics etc, where the multifractal analysis is also very successful. In particular, it can be used in data mining of financial data. The numerous studies (see, e.g. [10, 11, 12]) have shown that financial time series possess the mulifractal properties and hence the Holder exponents can be calculated. For the readers convenience we give below the basic information from the multifractal analysis.

### 2.1. Fractal dimension

In the definition of fractals, the main idea by Mandelbrot [13] has been related with the notion of selfsemilarity. In other words fractals are scale invariant objects. This scale invariance may be exact or statistical. It can be shown [13] that for fractals object the following relation is valid:

$$N(\varepsilon) \sim \varepsilon^{-D_F},$$

where $N(\varepsilon)$ is the number of cubes of the size $\varepsilon$ which form the covering of the set in question. The quantity $D_F$ is called the fractal dimension of the set. The power law dependence above leads to the scale invariance. The fractal set which is characterized by the unique universal parameter $D_F$ is called as monofractal [14].

One can define multifractal set as not uniform fractal object. It is obviously that for the whole description of multifractals it is insufficient to use the box-counting dimension $D_F$. It is necessary to introduce [15] the infinite number of the so-called generalized fractal dimensions (see, Section 2.2). Multifractal mathematical formalism exists in several versions [14-17]. Two approaches are briefly described below.

### 2.2. Generalized fractal dimensions

As in the case of fractal dimension definition let us use the box-counting method [14]. Suppose that the set in question is covered by cubes of size $\varepsilon$ and to each cube one can assign the quantity $\mu_i(\varepsilon)$ often called "mass". For example, if the set consists of isolated points the mass $\mu_i(\varepsilon)$ is the ratio of number of points in the cube with the number $i$ to the whole number of points in the set. So masses can be treated as probabilities that the point belongs to the given cube and describe relative occupancy of cubes from the covering. In case of homogeneous fractal, all cubes from the covering will contain the same number of points and hence all cubes will have the same relative occupancy. For continuum sets and multifractal measures the "masses" $\mu_i(\varepsilon)$ can be naturally introduced in each concrete case. The normalization condition $\sum_{i=1}^{N(\varepsilon)} \mu_i(\varepsilon) = 1$ is always valid. Here $N(\varepsilon)$ is the number cubes from the covering of the set in question. The so-called partition function or statistical sum which is characterized by the real parameter $q$ is introduced as follows:

$$Z(q,\varepsilon) = \sum_{i=1}^{N(\varepsilon)} \mu_i(\varepsilon)^q, \ q \in (-\infty, +\infty).$$

The scaling function is defined by the relation:

$$\tau(q) = \lim_{\varepsilon \to 0} \frac{\ln(Z(q,\varepsilon))}{\ln \varepsilon}.$$

In terms of these quantities, the generalized fractal dimensions $D_q$ are introduced by the equation [18]:

$$D_q = \frac{\tau(q)}{q-1} = \lim_{\varepsilon \to 0} \frac{1}{q-1} \cdot \frac{\ln\left(\sum_{i=1}^{N(\varepsilon)} \mu_i(\varepsilon)^q\right)}{\ln \varepsilon}.$$

It can be shown [17], that if for the given set the generalized fractal dimensions $D_q$ do not depend on $q$, then the set is monofractal. The scaling function $\tau(q)$ for monofractal is obviously linear.

### 2.3. Multifractal spectrum: Legendre transforms

Sometimes it is more convenient to use variables which are different from $D_q$ and $q$, but are closely related with the latter. Namely let us introduce the spectrum of multifractal dimensions $f(\alpha)$ where the variable $\alpha$ is the Holder-Lipschitz exponent for multifractal set in question. The change of variables from $\tau(q)$ and $q$ to $f(\alpha)$ and $\alpha$ is given by the Legendre transforms:

$$\alpha = \frac{d\tau}{dq}, \ f(\alpha) = q\frac{d\tau}{dq} - \tau(q),$$

$$q = \frac{df}{d\alpha}, \ \tau(q) = \alpha\frac{df}{d\alpha} - f(\alpha).$$

For multifractal objects, the condition $\frac{d^2\tau}{dq^2} \neq 0$ providing the existence of the Legendre transform is valid. One can show [18], that the quantity $f(\alpha)$ is equal in fact to the Hausdorff dimension of some homogeneous fractal subset which gives the principal contribution into the statistical sum at the given $q$. Since the fractal dimension of subset is always equal or less then the fractal dimension of the whole set it leads to inequality $f(\alpha) \leq D_0$. One can also show [14], that the function $f(\alpha)$ is convex for any multifractal set. The spectrum of multifractal dimensions $f(\alpha)$ is called also as the Legendre spectrum.

There is another approach [16, 17] to the spectrum of multifractal dimensions definition. Namely, consider a measure, which is distributed on the interval in arbitrary manner. Let us divide the support of the measure into equal boxes $C_\delta$ of the size $\delta$. Let $C = \{C_\delta\}$ be the set of all such boxes. Let us assume that inside of each box the measure satisfies to the relation $\mu\{C_\delta\} \approx \delta^{\alpha(C_\delta)}$, where the quantities

$$\alpha(C_\delta) = \frac{\log \mu(C_\delta)}{\log \delta}$$

are the so-called coarse-grained Holder exponents of the measure singularities. Let $N_\delta(\alpha,\varepsilon) = \#\{C_\delta : \alpha(C_\delta) \in (\alpha-\varepsilon, \alpha+\varepsilon)\}$, here (#) is the number of non empty boxes which measure is characterized by exponents from the interval $(\alpha-\varepsilon, \alpha+\varepsilon)$. Then the coarse-grained multifractal spectrum of large deviation [17, 18] is given by equation

$$f_g(\alpha) = \lim_{\varepsilon \to 0} \lim_{\delta \to 0} \sup \frac{\log N_\delta(\alpha,\varepsilon)}{\log 1/\delta}.$$

The last expression coincides formally with the definition of the capacity of a set if this set consists from boxes with the fixed value of the measure singularity exponent. The term coarse-grained is related with the finite precision of estimation of $\alpha$. The definition of $f_g(\alpha)$ is connected with the large deviation theorem which gives the probabilistic interpretation for the multifractal spectrum. Namely, the probability to find that $\alpha(C_\delta) \approx \alpha$ behaviors at small $\delta$ as follows:

$$N_\delta(\alpha,\varepsilon)/N_\delta = P_\delta[\alpha(C_\delta) \approx \alpha] \approx \delta^{d_\tau - f_g(\alpha)}.$$

Here $N_\delta$ is the total number of boxes covering the support of the measure and $d_\tau$ is the topological dimension of the measure support. Thus, the pair $\{\alpha, f_g(\alpha)\}$ describes the structure of an arbitrary measure. If the measure is multifractal, then its spectrum $f_g(\alpha)$ is a convex curve.

## 3. THE HOLDER EXPONENTS DEFINITION AND CALCULATIONS

As it was mentioned above, there are several approaches to the Holder exponents definition. For various reasons, these methods are not very convenient for numerical implementation. Therefore, we propose to use a method of modified Holder exponents, developed recently in [19]. For this purpose, we use the approach to the definition of the Holder exponents which is based on the introduction of so-called Holder semi-norm [19]. This approach is in short described below.

Definition 1. *Function $f(x)$, defined on the domain $E \subset R$ satisfies on the set $E$ the Holder condition with the exponent $\alpha$, where $0 < \alpha \leq 1$, and with the coefficient $A$, if*

$$|f(x) - f(y)| \leq A|x-y|^\alpha$$

*for any* $x, y \in E$.

The quantity

$$C_a = \sup_{x,y \in E} \frac{|f(x) - f(y)|}{|x-y|^\alpha}, \quad 0 \leq \alpha \leq 1, \tag{3.1}$$

is called the Holder $\alpha$-semi-norm of the bounded function defined on the set $E$.

Let $f(x)$ satisfies the Holder condition with the exponent $\alpha$ on the set $E$. Then one can easily show that

$$\begin{cases} C_\beta = 0, \text{ if } \beta < \alpha, \\ 0 < C_\beta < +\infty, \text{ if } \beta = \alpha, \\ C_\beta = +\infty, \text{ if } \beta > \alpha. \end{cases}$$

The behavior of $C_\beta$ depending on $\beta$ is shown in Fig. 3.1 by the dotted line.

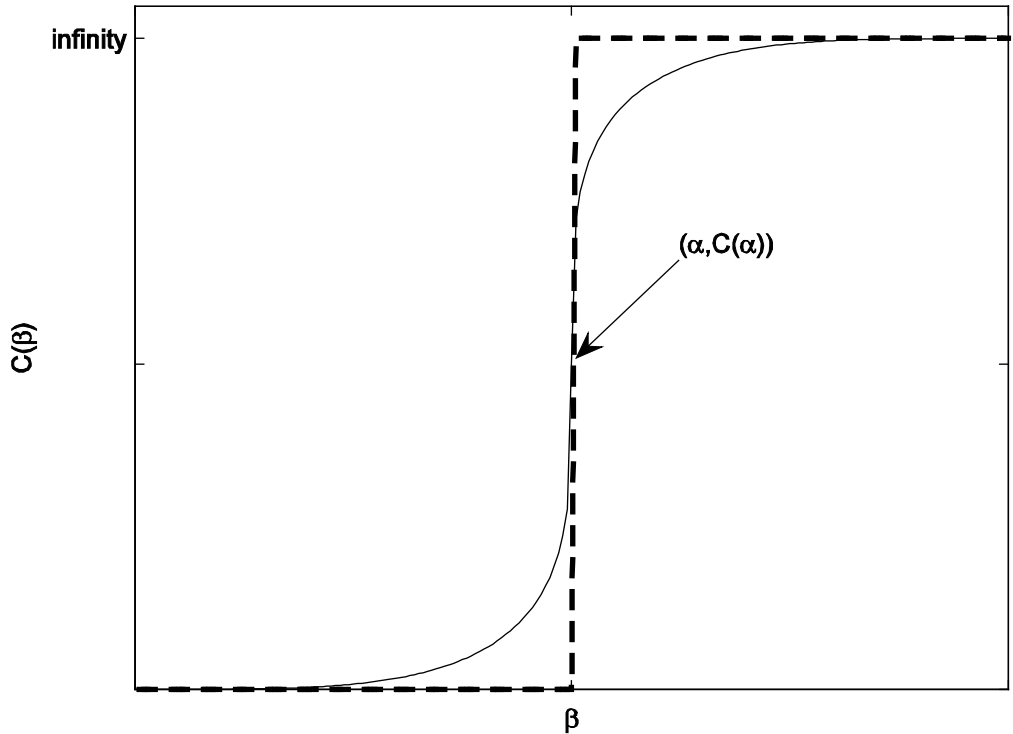

Fig. 3.1. The behavior of $C_\beta$ versus $\beta$ in continuous (dotted line) and discrete (solid line) cases.

From the definition of $C_\beta$ it follows that there is a unique value of $\beta$, such that

$$0 < C_\beta < +\infty, \tag{3.2}$$

and such $\beta$ coincides with the Holder exponent $\alpha$. This property gives the algorithm of the Holder exponent $\alpha$ calculations. In accordance with equation (1) one calculates $C_\beta$ for all $0 \leq \beta \leq 1$. Further the value of the coefficient $\beta$, which satisfies the condition (2) is chosen. This value of $\beta$ is the required Holder exponent $\alpha$. It is obvious that numerical realization of the algorithm described above is embarrassed by difficulties connected with the control of the right hand side of condition (2). Following [19] the numerical realization of the algorithm for time series is described below.

Let variable $t$ takes the discrete values $t_i$ from the interval [0,1]: $t_i = i/N$, $i = \overline{1..N}$. Let $X_i$ be the value of function $f$ at points $t_i$. In that way one obtains the discrete time series $\{X_i\}_{i=1}^{N}$. For the algorithm realization, it is also necessary to define the increment of the parameter $\beta$. Toward this end let us fix the natural number $n$, define the increment as $\Delta\beta = 1/n$ and introduce the quantities $\beta_k = k/n$, where $k = \overline{0..n}$. In the discrete case, the equation (1) takes the form:

$$C(\beta) = \max_{i,j=1..N, i \neq j} \frac{|X_i - X_j|}{|t_i - t_j|^\beta}, \quad \beta = \beta_k, k = \overline{1..n}. \tag{3.3}$$

One has to study the behavior of $C(\beta_k)$ for different $k$. For that let us introduce the value

$$\overline{k} : |\alpha - \beta_{\overline{k}}| = \min_{k=1..n} |\alpha - \beta_k|,$$

where $\alpha$ is the Holder exponent of the function $f$. It means that $\beta_{\overline{k}}$ is the best approximation of $\alpha$ among all $\beta_k$. Expressing $C(\beta_k)$ in terms of $C(\beta_{\overline{k}})$ one obtains

$$C(\beta_k) = C(\beta_{\overline{k}}) \max_{i,j=1..N, i \neq j} \frac{1}{|t_i - t_j|^{\beta_k - \beta_{\overline{k}}}}.$$

If $k < \overline{k}$, that is $\beta_k < \beta_{\overline{k}}$, then for sufficiently large $n$ and $N$ the value of

$$\max_{i,j=1..N, i \neq j} \frac{1}{|t_i - t_j|^{\beta_k - \beta_{\overline{k}}}},$$

and hence the value of $C(\beta_k)$ will be close to zero. The smaller the value of $k$ one takes, the closer to zero the quantity $C(\beta_k)$ becomes. In a similar manner at $k > \overline{k}$ the value of $C(\beta_k)$ will be much greater then 1 for sufficiently large $n$ and $N$. The quantity $C(\beta_k)$ grows with $k$ increasing. While passing from $k = \overline{k} - 1$ to $k = \overline{k} + 1$ the function $C(\beta_k)$ has the sharp jump. Summing the information about $C(\beta_k)$ behavior, one can obtain the plot of $C = C(\beta_k)$ (Fig 1, solid line). In the discrete case, the obtained curve $C(\beta_k)$ is an approximation of that in the continuous case. The greater $n$ and $N$ values one takes, the more precise approximation is obtained.

In order to investigate the multifractal properties of time series generated by a modified Minority Game we need proposed in [19] modified Holder exponent (MHE) and modified point wise Holder exponent (MPHE). Method of their construction are summarized below. Let us consider the first $N'$ values $\{X_i\}_{i=1}^{N'}$ of the time series $\{X_i\}_{i=1}^{N}$. Suppose that for the time series $\{X_i\}_{i=1}^{N'}$ the Holder exponent is equal to $\alpha$. Knowing $\alpha$, let us calculate $C(\alpha)$ for the set $\{X_i\}_{i=1}^{N'}$ by means of the equation (3). Further for the set $\{X_i\}_{i=N'+1}^{N}$ the quantities $C(\beta_k)$ are calculated. But the procedure of calculations will be changed. In step by step calculating the numbers $C(\beta_k)$ let us stop at that moment when $C(\beta_k)$ runs up to the value of $C(\alpha)$. Let it happens at the time $\overline{k}$. Then one defines the quantity $\overline{a} = \beta_{\overline{k}}$. Thus for arbitrary $\alpha$ one can calculate $\overline{\alpha} = \overline{\alpha}(\alpha)$. The indicator $\overline{a}$ will be called the modified Holder exponent (MHE) of the set $\{X_i\}_{i=N'+1}^{N}$ with respect to the set $\{X_i\}_{i=1}^{N'}$.

For the purpose of our study let us remind [20] the notion of the Holder exponent at some point $t_0$.

**Definition 2.** *Function $f(t)$ has the Holder exponent $\alpha$ at the point $t_0$, iff*
1. *for any real $\gamma < \alpha$*

$$\lim_{h \to 0} \frac{|f(t_0 + h) - P(h)|}{|h|^\gamma} = 0$$

*and*

2. *if $\alpha < \infty$, for any real $\gamma > \alpha$*

$$\limsup_{h \to 0} \frac{|f(t_0 + h) - P(h)|}{|h|^\gamma} = +\infty,$$

*where $P$ is a polynomial of order not greater then $\alpha$.*

Following [19] we shall call the MHE calculated at some point as the point wise MHE (MPHE). The values of MPHE have been normalized since at the points of local maximum their values were greater then 1. However, the character of increasing and decreasing of the Holder exponents and of MPHE looks very similar.

## 4. STASTATISTICAL CHARACTERISTICS OF FINANCIAL TIME SERIES

### 4.1. Stable probability distributions

Consider a few basic differences in the distribution of real data from a normal distribution. The first is the presence of both "fat tails" - for large values of standard deviations of the distribution function decays slower than exponential, namely, according to power law. Also the distribution of real data is more acute than in the normal distribution peaks. As an alternative to the normal distribution B. Mandelbrot proposed the use of the Levy distribution. The characteristic function [21] of Levy stable distributions is given by:

$$\ln F(q) = i \cdot \delta \cdot q - c \cdot |q|^\alpha \cdot \left[1 - i \cdot \beta \cdot \frac{q}{|q|} \cdot tg \frac{\pi \alpha}{2}\right], \alpha \neq 1$$

$$\ln F(q) = i \cdot \delta \cdot q - c \cdot |q| \cdot \left[1 - i \cdot \beta \cdot \frac{q}{|q|} \cdot \ln|q|\right], \alpha = 1$$

Distribution of Levy is characterized by four parameters:
$\alpha \in [0, 2]$
$\beta \in [-1, 1]$
$c \in R^+$
$\delta \in R$

Here $\alpha$ is a certain parameter often called characteristic exponent, $\beta$ is the skewness parameter, $c$ is the scale parameter and $\delta$ is the location parameter. Characteristic exponent is very important. The bigger value it have the fatter distribution function tails. The skewness depends on scale parameter. In some papers [22] scale parameter is called standard deviation, but it is wrong because of for $\alpha < 2$ there is no standard deviation and for $\alpha = 2$ variance is equal $2c^2$. Likewise, the location parameter is not the mean, unless $\alpha > 1$ и $\beta = 0$.

However, later researches have provided substantial evidence that empirical distribution tails drop-down faster than Levy distribution tails [17]. Moreover, as a rule, real date have finite time dependence variance. Stability can be observed only on the little time scale, in the other cases convergence to normal distributions is observed.

To overcome this problem P. Mantegna and H. Stanley [23, 24, 25] proposed truncated Levy distribution for real date. In case of truncated Levy distribution central part is described by stable

Levy and tails have exponential drop-down. Such modification provides Gaussian convergence and finite variance.

To date, there are many moot points concerning probability distribution function of real data. Nevertheless, this is more appropriate version with mathematical point of view that good describe the real data.

## 4.2 R/S Analysis

Method of R/S analysis or rescaled range analysis is one of the oldest and best-known methods. Bellow algorithm of estimating Hurst exponent by this method is adduced.

1. Split the series (series of logarithmic returns) $r_i$, $i=1...L$ length $L$ into $d$ bins size $n$. Then every element in bins will be $r_{j,k}$, $j=1...n$, $k=1...d$.

2. Then for every bins of series $k=1,......d$, compute mean ($E_k$) and standard deviation ($S_{k,n}$).

3. After that cumulative sum should be computed

$$Y_{m,k} = \sum_{j=1}^{m}(r_{j,k} - E_k), \quad m=1,......n \tag{4.1}$$

And find so-called range

$$R_{k,n} = \max Y_{m,k} - \min Y_{m,k}. \tag{4.2}$$

4. Find relation

$$<R/S>(n) = \frac{1}{d}\sum_{k=1}^{d}\frac{R_{k,n}}{S_{k,n}}. \tag{4.3}$$

5. As a definition process is called Hurst process if:

$$<R/S> \propto n^H, \text{ or } \log<R/S> = H \cdot \log n + const \tag{4.4}$$

As the output of this algorithm, we have the logarithms of the intervals lengths and the logarithms of the ratios (R / S) for each interval. Plotted logarithms (R / S) as a function of the logarithms of the intervals lengths and the approximating line, we obtain the Hurst exponent (slope of approximating line). The errors of the algorithm are calculated as the ratio of the amount of all deviations to the length of time series.

## 4.3 Volatility clustering

It is well known that the real financial time series have the property of volatility clustering. Under the volatility in general we mean measure of the change in time of a certain discrete values, in this case - increment of market price [6]. Some result for the real time series is shown in Fig. 4.1.

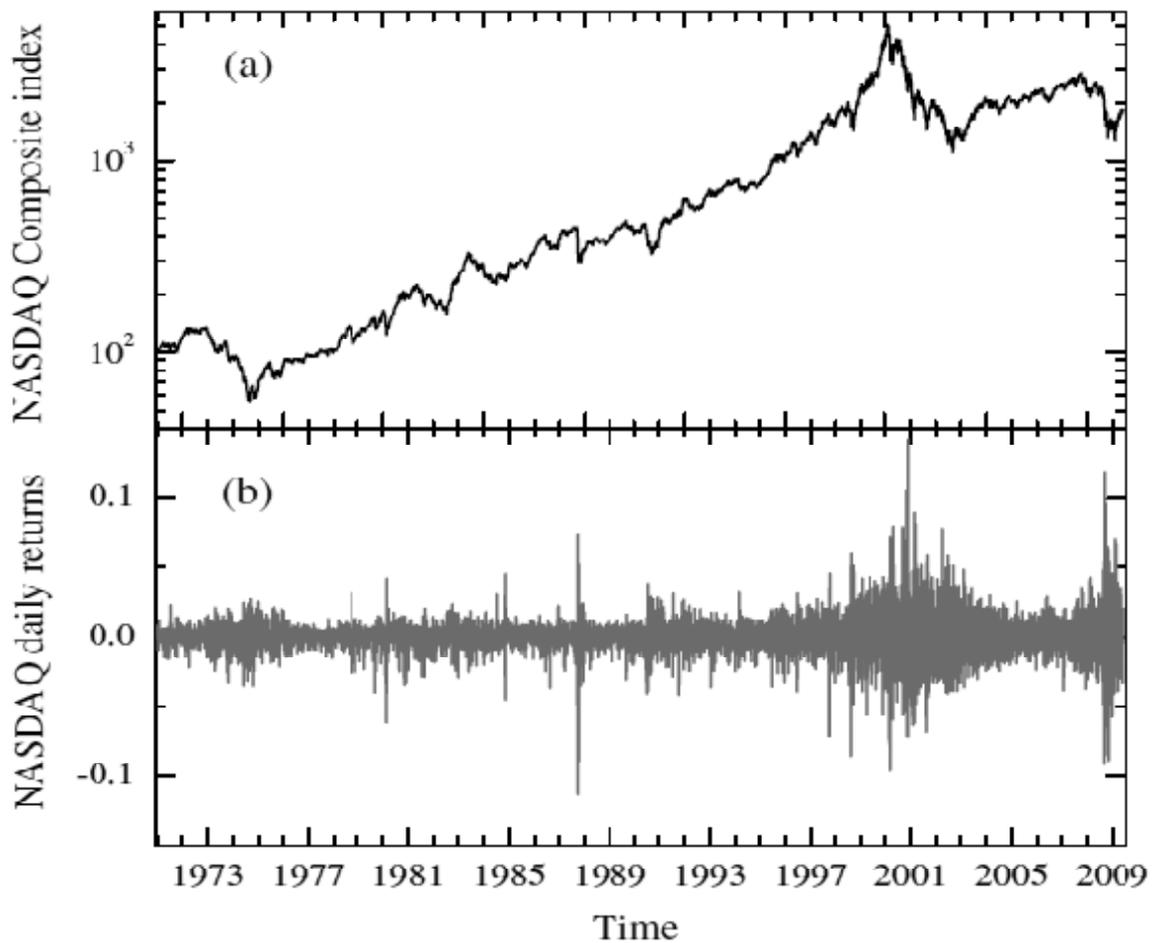

Fig. 4.1. Empirical evidence NASDAQ index for the period from 8 February 1971 to 30 June 2009 with a daily slice. Panel (a) shows the daily closing price, and panel (b) - schedule of daily increments during this period [6]. The bottom panel clearly shows the effect of volatility clustering.

There are many other definitions of volatility, but we will adhere to this, because it allows fairly easy to define the following key concept - clustering of volatility. Under the volatility clustering effect imply grouping of both large and small fluctuations in the increment of prices of financial instruments [6]. As has been noted by B. Mandelbrot, "for a large change, usually follow other large changes of any sign, but for small changes, usually follow small changes" [26] (see Figure 2).

A more precise definition proposed by Cont [27]. Although the price changes by themselves do not correlate the sequence of their absolute values or squares have a positive, separated from zero and slowly decaying autocorrelation function. For a quantitative measure of volatility was developed several algorithms. For example, Cont [27] proposed evaluation method for clustering of volatility, linking it to the investors inertness. He developed a model of changes in market price of an asset which can not only generate a sequence of price changes over time, but also to evaluate clustering of volatility. During the period of low volatility of prices the investor's inertness is large, which explains the clustering of low volatility. During the period of high volatility of prices increases the activity of agents, which leads to clustering of high volatility.

Despite the convenience of calculating volatility clustering by the method of Cont, in this paper was applied a different method [6], which seems to be more straightforward from the computational point of view. In the method proposed by Tseng Jie-Jun [6], volatility clustering coefficient $R$ is calculated by comparing the characteristics of changes in real prices and the characteristics of the sequence whose values are distributed by Gaussian. Namely, introduce the clustering coefficient by the equation:

$$R \equiv \sigma_e / \sigma_G \qquad (4.5)$$

Here $R$ is the factor clustering, $\sigma_e$ is the standard deviation of p% of the most significant normalized fluctuations studied sequence, and $\sigma_G$ is the standard deviation of p% of the most significant fluctuations in the sequence, distributed by the Gaussian law.

Let us consider the mechanism for calculating $\sigma_e$ and $\sigma_G$. Namely, from the test range of prices $p_t$ calculated time series of returns $R_i$. This time series $R_i$ is subjected to normalization, which is bringing this series to the form in which the average is 0 and standard deviation is 1. This process is described by (4.6):

$$\{E_i\}^N_{i=1} : E_i = \frac{R_i - \mu_R}{\sigma_R} \qquad (4.6)$$

Then the resulting time-series and series distributed by Gaussian, sorted by descending, as shown in formulas (4.5) and (4.6), where $\mu_R$ is the expectation, and $\sigma_R$ is the standard deviation:

$$\{i_k\}^N_{k=1} : E_{i1} \geq E_{i2} \geq ... \geq E_{iN} \qquad (4.7)$$

$$\{j_l\}^N_{l=1} : G_{j1} \geq G_{j2} \geq ... \geq G_{jN} \qquad (4.8)$$

where $G_{jk}$ is the time series, distributed by Gaussian. For further processing from the beginning of sorted series takes a certain percentage (p%) samples from the sorted series:

$$M = N \cdot p\% / 100\% \qquad (4.9)$$

In the formula (4.9) M is the number of samples, which we take from the series for research, and N is total number of samples in these series. As can be seen, M depends on the p%, which is chosen in a special way in each case. Then the standard deviations of the series are substituted into the formula (4.5) and calculate the clustering coefficient. According to [6] if the clustering coefficient is greater than unity, this means that in the investigated time series is an effect of clustering of volatility.

## 5. MODIFICATIONS OF THE MINORITY GAME

As stated above, all the major modifications made to the classical model of Minority Game have to meet the requirement of realistic games. In other words, the goal is to bring modifications introduced by the modified Minority Game to the realities of the market. Below we describe these modifications. First, unlike the original version of Minority Game each player was assigned a start-up capital, which they can lose or increase in future games. This was possible because of the availability of rates, which became the second significant change compared to the classical version of the Minority Game. Now every agent in each of his decision at each stage of the game puts a certain part of their capital, which is a function of capital itself, the number of its zero strategies (this concept of zero strategies will be explained below) and the total number of strategies. The studies, whose results are presented in this paper, everywhere used the same formula for the rate of players:

$$\frac{f}{3s} b \qquad (5.1)$$

In the formula (5.1) $f$ is the number "zero" strategy of the player, $s$ is the total number of strategies, and $b$ is the current balance of the player. Thus, depending on how many strategies require a player simply pass play (risk-taking or risk-free strategy), the player can put a third of its capital or do not put anything. Payments made by the following rule: if the player happened to be in the minority, he got his bet double the amount, if turned out in most - losing bet.

The definition of zero strategy is: a third choice behavior allowed for each player in this modification Minority Game. If a player does not see the benefits of participating in the game at the moment, he could miss the game. In order that in the end all strategies are not nullify (that is, in principle, possible, because taking no part in the game the player loses nothing), the game has the possibility of small fluctuations in the strategies pips.

The price of an abstract resource was calculated using the following formula (5.2):

$$P_i = P_{i-1} - B + A \qquad (5.2)$$

Here $P_i$ is the price of $i-th$ stage of the game, $B$ is the number of selling players (those who chose the strategy of "-1"), $A$ is the number of buying players (those who chose the strategy of "1"). Players who did not put the game out of the total counts are excluded. Also in the calculation does not include players who had lost all their money. They do not call the game, because there is no opportunity to borrow capital. However, as will be shown below, even such a simple calculation of prices for an abstract resource allows obtaining the desired effects for the time series returns of players.

The last major modification was the introduction of rotating the players' strategies. We define the concept of the strategies rotation. Namely, if a strategy is no longer useful to the player (the number of point's strategy becomes zero), it is replaced not by a random strategy, but by a strictly defined. For example, in the present work, the players, who bought up, pass game; selling players also miss the game; but those who missed the game, go buying. In scores of zero strategies introduced fluctuations, which do not allow all players to miss the game. Namely, at each step of the zero points, strategies randomly subtracted number in the range from 0 to 2. It will be shown that rotation of the strategies leads to multifractality of time series of prices for an abstract resource.

## 6. NUMERICAL RESULTS AND THEIR ANALYSIS

### 6.1 Statistical tests

First, using package STATISTICA on the level of confidence = 0,05 was considered the hypothesis of normal distribution of logarithmic returns obtained using the modified Minority Game time series of prices. In Figure 6.1. the distribution of logarithmic returns of the time series from 1000 price values generated by the game for 500 players with memory 10 is represented. At a significance level of 0.05 critical value of $\chi^2$ statistics equal to 18.31. As seen from the graph the observed value of $\chi^2$ statistics equal to 252.42, while the corresponding value p-level virtually equal to 0. Therefore, according to the $\chi^2$ test the null hypothesis of normality of distribution should be rejected.

Figure 6.2 shows the distribution of logarithmic returns of the time series of 5500 values generated by the game with number of player = 800 with the value of memory = 12. At a significance level = 0.05 critical value of $\chi^2$ statistics equal to 12.59. As seen from the graph, the observed value of $\chi^2$ statistics equal to 1278, and the corresponding value p-level virtually is equal to 0. Therefore, according to the $\chi^2$ test the null hypothesis of normal distribution also should be rejected. Thus, the hypothesis of normal distribution of logarithmic returns series was rejected.

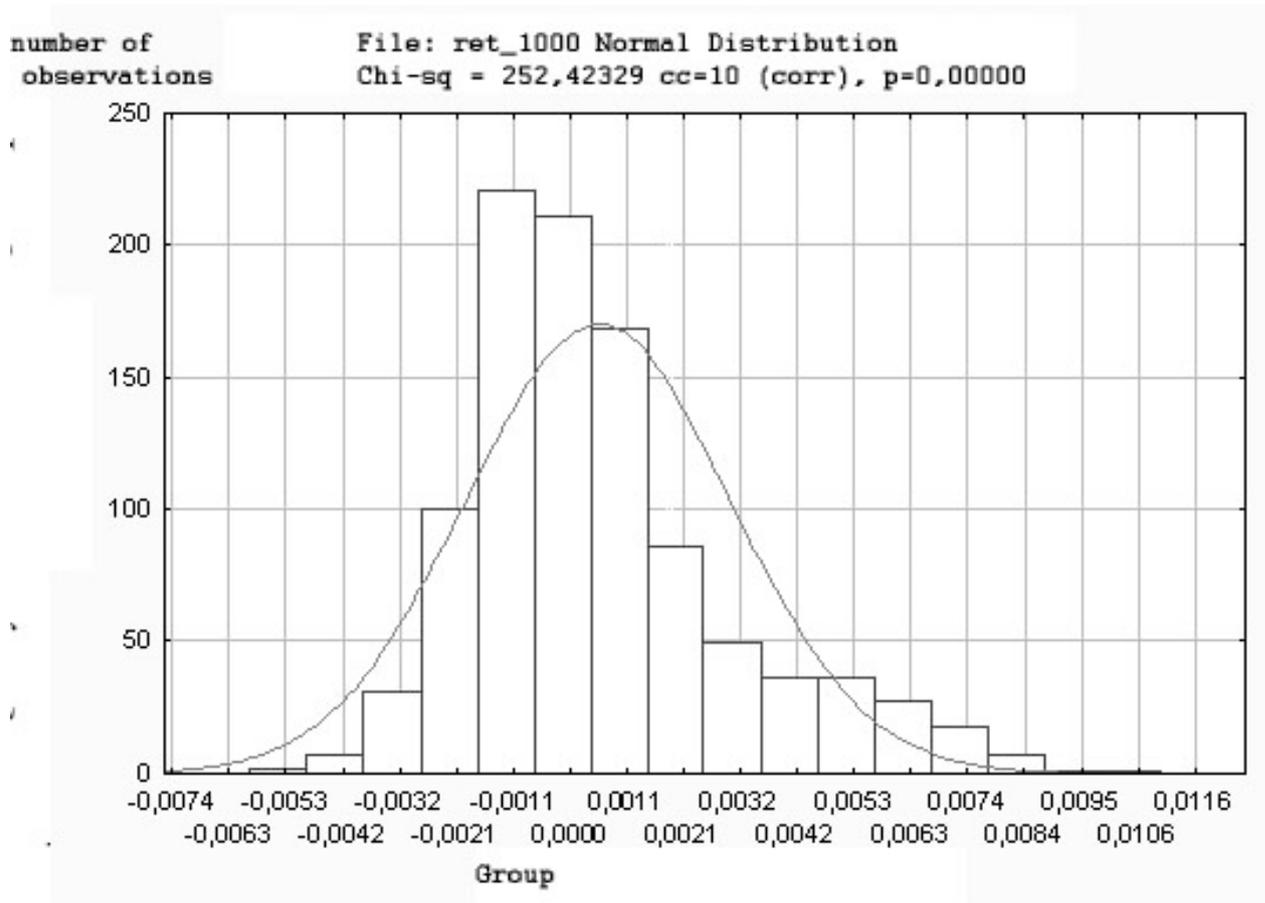

Figure 6.1. Approximation by the normal distribution of a number of logarithmic returns in 1000 values generated by the game of 500 agents with the value of memory equal to 10.

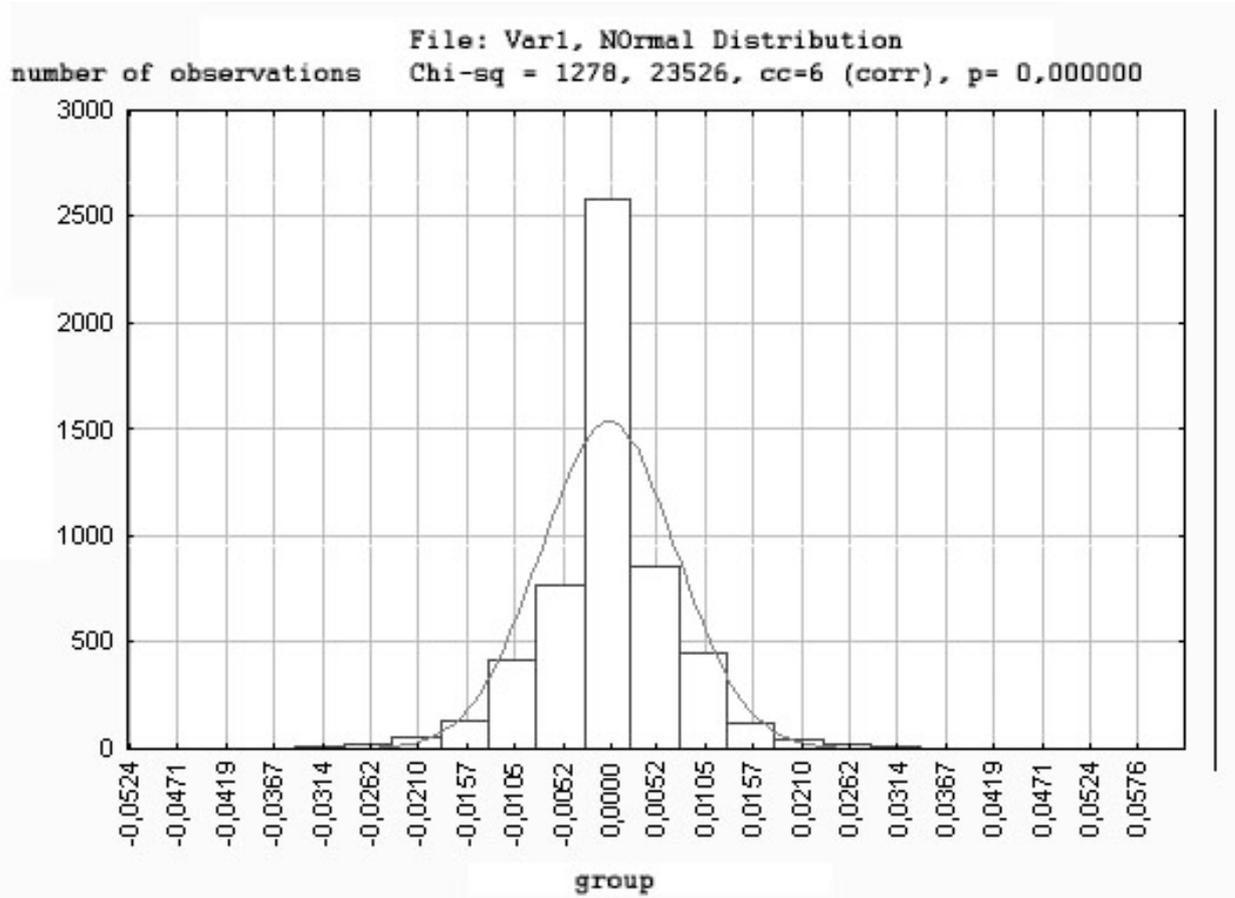

Figure 6.2. Approximation of the normal distribution of a number of logarithmic returns of the 5500 values generated by the game of 500 agents with a value of memory of equal to 12.

After refuting the hypothesis of normal distribution there has been tested the hypothesis of distribution of data according to the Levy distribution. This was done in the module "Stable motion" of the package Fraclab [28]. The results of the parameters fitting of the Levy distribution presented in Figure 6.3.

| | Outputs | |
|---|---|---|
| | Value | std |
| Characteristic Exponent | 1.4647 | 0.08521 |
| Skewness Parameter | 0.85601 | 0.16608 |
| Location Parameter | 0.00097898 | 0.00023287 |
| Scale Parameter | 0.001268 | 5.3514e-005 |

Figure 6.3. The results of processing the time series of 1000 values (100 players with memory 10) using the module "Stable Motion" of the Fraclab package [28].

The results of fitting the parameters of the Levy distribution for different number of players and different memory are presented in Figure 6.4.

| | Outputs | |
|---|---|---|
| | Value | std |
| Characteristic Exponent | 1.2429 | 0.022508 |
| Skewness Parameter | -0.045781 | 0.04039 |
| Location Parameter | -0.00020078 | 0.00026886 |
| Scale Parameter | 0.0029262 | 5.5504e-005 |

Figure 6.4. The results of processing the time series of 5500 values (800 players with memory 12) using the module "Stable Motion" the Fraclab package [28].

Thus, the hypothesis of distribution of data obtained using the modified Minority Game, according to the Levy distribution confirmed.

### 6.2 Multifractality tests

Before we check the time series on multifractality, one should check whether they can all be fractal. To do this, we calculated the Hurst exponent [14] by R/S analysis, described in Section 4.2 of this paper.

```
Final Hurst exponent is 0.623237
Constant C calcualted as -0.426207
Calculation error is 0.000858
```

Figure 6.5. Computation of the Hurst exponent for a time series of 1000 values of prices generated by the game 500 players with memory equal to 10.

Figure 6.5 shows the result of the program that implements the appropriate algorithm. The value of Hurst exponent H = 0,6232 ± 0,0009. The error was calculated as the ratio of the sum of all deviations to the length of time series. Thus, within the error of Hurst index exceeding 0.5. The constant in the formula (4.4) C = -0,4262. Based on the value obtained Hurst exponent, it is clear that the resulting time series is a fractal.

We were also studying time series of the modified Minority Game on multifractality. It has been done using a special program written in Matlab package. Namely there has been obtained a series of modified local Holder exponents (MPHE) [19], whose calculation is described in section 3 of this work. The behavior of MPHE is shown in Figure 6.6. The graph shows that a number of exponent's samples is constant, but then their values begin to vary. This suggests that the studied time series is multifractal everywhere except in those first readings MPHE. Thus, the hypothesis that the time series generated by the proposed version of Minority Game are multifractals, was confirmed.

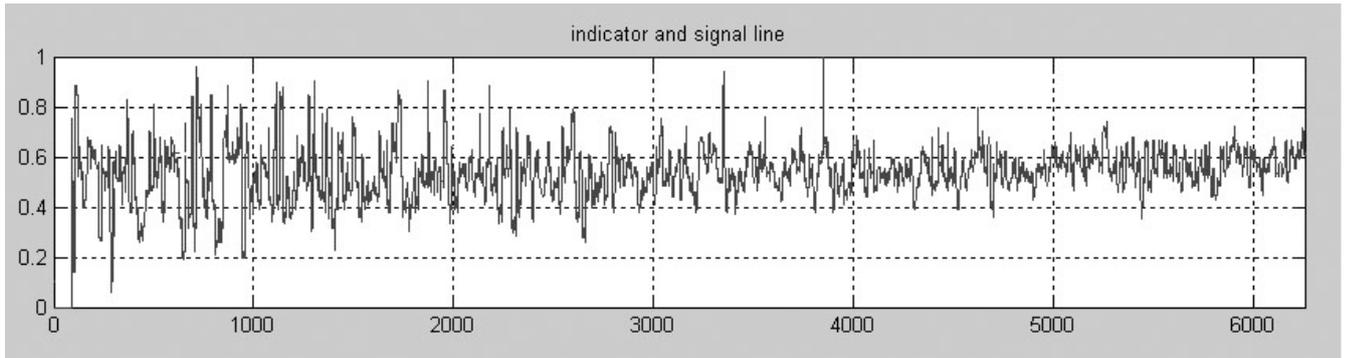

Figure 6.6. The graph of MPHE for a number of the 1000 values, generated by the activity of 500 players with memory equal to 10. The ordinate values of the MPHE exponents and the abscissa time readouts exponents.

**6.2 Volatility clustering**

Availability check of the volatility clustering in time series generated by a modified Minority Game was performed by the method described in the section 4 of this work. This algorithm has been implemented through the program written for the purpose. The result of the program operation is a table, shown in Figure 6.7.

```
Read 1000 numbers from 'e.txt'
Read 1000 numbers from 'g.txt'
deviation for the first series is 9.256879
deviation for the second series is 7.074848

clustering coefficient is 1.308421
```

Figure 6.7. The results of the program operation for the calculation of volatility clustering for the logarithmic returns of a number of 1000 samples generated by the activity of 500 players with the memory 10.

As can be seen from figure 6.7, standard deviation for the test series $\sigma_e = 9.2569$, and the standard deviation for time series, distributed according to Gauss $\sigma_G = 7.0748$. Clustering coefficient, the ratio of these quantities, has the value $R = 1.3084$. Clustering coefficient significantly exceeds unity, which indicates the presence of volatility clustering for time series of prices of the modified Minority Game. Visualization of the effect of volatility clustering for time series generated by a modified Minority Game presented in Figure 6.8.

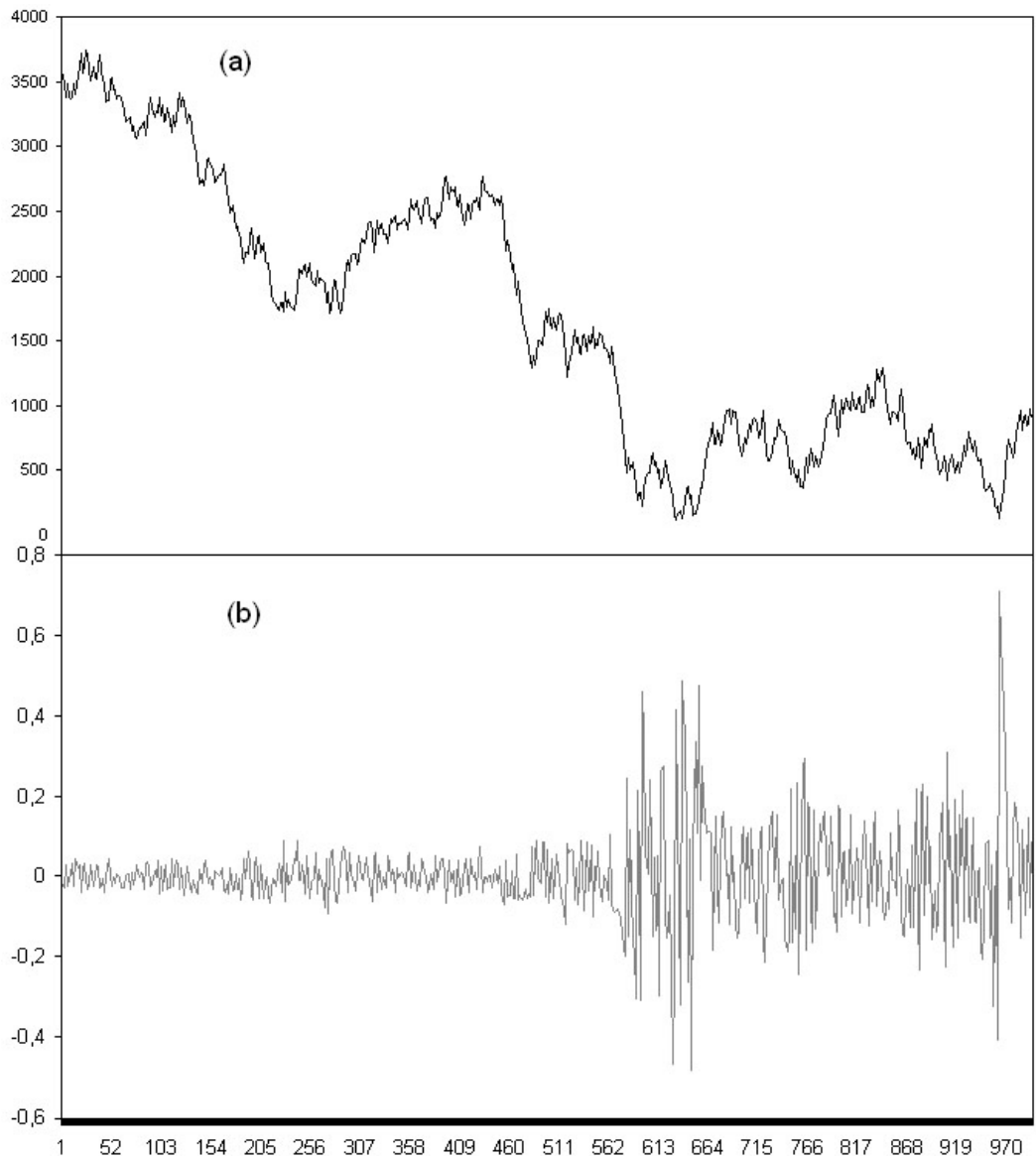

Figure 6.8. Time series of 1000 time samples of the modified Minority Game. Figure (a) shows the price chart, and figure (b) shows graph of returns. Figure 6.8 (b) can also clearly distinguish the presence of clusters.

**SUMMARY**

Summarize the results which were obtained in this study. First, a modification of the known multiagent model Minority Game was developed and designed to simulate the behavior of traders in financial markets and the resulting price dynamics on the abstract resource. Secondly, this model was implemented in the form of software. Thirdly, the proposed model was explored for reproducing the basic properties of real financial time series. It was proved that such properties as the clustering of volatility, the Levy distribution and inherent multifractality of time series, can be generated by the proposed model.

Besides, another important result was obtained, as described below. In all works designed to simulate the property of multifractality of financial series, researchers in order to get the result had to go to steps which seems to be not quite realistic. Most of the players were divided into several groups, who were allowed to trade only in different time periods. Certain groups of agents have been denied in the rational behavior. In the present study, for obtaining multifractality another step has been taken. This step is introduction into a model a new trick: the strategies rotation. Strategies rotation is a mechanism for changing the strategy with nullified points of total utility. If in previous studies strategies were changed randomly, in the present paper they are changed by a well defined rule. This completely meets the demands of realism. After all, if the trader changes his strategy, then this step is usually due to its low-value utility and has the aim to increase the utility value. Thus, this work also suggests a new mechanism of generation of multifractal time series which is more realistic than in the above-mentioned works.

However, this study leaves several open issues that are of interest for future studies. Among them is, for example, the study of waste, which can be made by the traders population depending on the conditions of the game (this self-organized aspect was considered in the paper [29]). It is also interesting to give a qualitative explanation of the causes of the obtained time series multifractality. Also there is an open question about the regularities in the dynamics of the player capital.